\documentclass[twocolumn]{revtex4}
\usepackage{graphicx}
\usepackage{siunitx}
\usepackage{lineno}
\usepackage{amsmath}
\usepackage{comment}
\usepackage{color}

\def\dd{{\rm d}}

\usepackage[sort&compress]{natbib}
\begin{document}

\setcounter{page}{1} 

\title{Quantifying the Sensitivity of HIV-1 Viral Entry to Receptor
  and Coreceptor Expression} 

\author{Bhaven Mistry$^{1}$, Maria R. D'Orsogna$^{1,2}$, Nicholas
  E. Webb$^{3}$, Benhur Lee$^{4}$, and Tom Chou$^{1,5}$ \\ 
$^{1}$Department of Biomathematics, University of California, Los
  Angeles, CA 90095 \\ 
$^{2}$Department of Mathematics, California
  State University, Northridge, CA 91330 \\ 
$^{3}$Department of Infectious Disease, Children's Hospital Los Angeles, Los Angeles, CA
  90027 \\ 
$^{4}$Department of Microbiology, Icahn School of Medicine
  at Mount Sinai, New York, NY 10029 \\ 
$^{5}$Department of Mathematics, University of California, Los Angeles, CA 90095}

\email{tomchou@ucla.edu}
\begin{abstract}
Infection by many viruses begins with fusion of viral and cellular
lipid membranes, followed by entry of viral contents into the target
cell and ultimately, after many biochemical steps, integration of
viral DNA into that of the host cell.  The early steps of membrane
fusion and viral capsid entry are mediated by adsorption to the cell
surface, and receptor and coreceptor binding. HIV-1 specifically
targets CD4+ helper T-cells of the human immune system and binds to
the receptor CD4 and coreceptor CCR5 before fusion is
initiated. Previous experiments have been performed using a cell line
(293-Affinofile) in which the expression of CD4 and CCR5 concentration
were independently controlled.  After exposure to HIV-1 of various
strains, the resulting infectivity was measured through the fraction
of infected cells.  To design and evaluate the effectiveness of drug
therapies that target the inhibition of the entry processes, an
accurate functional relationship between the CD4/CCR5 concentrations
and infectivity is desired in order to more quantitatively analyze
experimental data.  We propose three kinetic models describing the
possible mechanistic processes involved in HIV entry and fit their
predictions to infectivity measurements, contrasting and comparing
different outcomes.  Our approach allows interpretation of the
clustering of infectivity of different strains of HIV-1 in the space
of mechanistic kinetic parameters. Our model fitting also allows
inference of nontrivial stoichiometries of receptor and coreceptor
binding and provides a framework through which to quantitatively
investigate the effectiveness of fusion inhibitors and neutralizing
antibodies.
\end{abstract}

\maketitle

\section{Introduction}
Despite their great adaptability and capacity to survive in many
different environments, viruses are not equipped with the necessary
biochemical materials, structures, or metabolic resources to self
replicate \cite{PMC3405824, Roche, opac-b1099993, Boulant,
  Stalmeijer15032004,Chou2,Nowak,Chou3}. In order for a virus strain
to survive, it must find and bind to a host cell membrane, inject its
virion contents (RNA, reverse transcriptase, proteins) into the
cytosol of the host cell through membrane fusion or endocytosis, help
facilitate the processing and transport of such contents to the cell
nucleus, and finally integrate its genome into the DNA of the host
cell. After these complex series of events, the ``hijacked'' cell is
instructed to produce the virion's constituent parts that later
assemble into new viruses and escape the host cell
\cite{opac-b1099993,Qian,PMC3405824, Boulant,Chou2,Nowak}. 

Therapies developed to combat viral infection involve inhibiting one
or several of the above described processes employed by the virus to
infect the target cell \cite{Qian,Pegu243ra88, nbt0898-778}. For
example, in the case of the human immunodeficiency virus (HIV),
enfuvirtide (T-20) inhibits fusion of the viral membrane with that of
the host cell \cite{Su_Qiu_2012} while zidovudine, didanosine, and
zalcitabine inhibit reverse transcription of RNA into
integration-ready DNA \cite{MED:MED2}. Elvitegravir, dolutegravir, and
raltegravir inhibit DNA integration in the nucleus, blocking the
insertion of the viral genome into the host
DNA \cite{thierry2015integrase}, while darunavir, saquinavir, and
fosamprenavir inhibit HIV-1 protease activity which ensures the proper
cleavage of viral polypeptide chains \cite{Paterson2000}. Fusion
inhibitors, the latest class of anti-viral drugs to be developed, are
now integrated into overall therapy and have the advantage of
inhibiting the virus at the first step of the infection process. 
In principle, the employment of fusion inhibitors can reduce the need for
intracellularly targeted therapies which often result in additional
side-effects, require higher drug concentrations, and involve more
complicated pharmacokinetics \cite{Su_Qiu_2012}.

In order to design and assess the effectiveness of viral entry
inhibitor treatments, a quantitative description of the viral entry
process is necessary \cite{Moore2003,Platt2005}. Since a complete
biological picture is still elusive, mathematical models that include
relevant mechanisms such as surface receptor binding, dissociation,
viral degradation, and viral fusion can allow us to explore several
aspects of the viral entry process, such as the influence of
multiplicity of infection and stoichiometry of receptor/coreceptor
binding.  Once appropriate models are derived, a statistical inference
can be performed using data from experimental measurements of viral
entry to validate assumptions and to obtain constraints on the model
rate parameters.
 
HIV-1 is an enveloped virus that follows a receptor-coreceptor binding
paradigm \cite{PMC3405824, opac-b1099993,Platt1,Nowak}. It has evolved
to target helper T-cells of the human immune system. Helper T-cells
express the membrane surface receptor CD4 which normally functions as
a coreceptor to the T-cell receptor (TCR) complex. Both the TCR and
CD4 bind to the MHC class II protein complex of antigen-presenting
cells (APC) prior to the initiation of cytokine release,
activation of cytotoxic T-cells and antibody producing B-cells, and
other secondary immune response processes \cite{opac-b1099993}.

HIV-1 binds to CD4 with its glycoprotein spike, Env, upon contact with
the cell surface of the helper
T-cell \cite{Boulant,PMC3405824,Platt1,Nowak}. Env is formed by a
trimer of a pair of glycoprotein subunits: gp120 and gp41. The former
subunit contains a CD4 binding domain and five variable chain
regions. The complex gp41 anchors gp120 to the viral membrane via a
non-covalent bond and is central to the fusion process of
infection \cite{Manca01091993, White2008}. After binding to CD4, gp120 goes
through a conformational change that exposes an occluded binding
domain that binds to a cell surface coreceptor. Though many
coreceptors have been identified and are used differentially depending
on the HIV-1 strain, the two that are by far the most prevalent are CCR5
and CXCR4 \cite{PMC3405824, Xiao1999}. HIV-1 that bind CCR5 are called R5 strains
and are the most common variant involved in transmission of the virus
between individuals while the strains that bind CXCR4, R5X4, normally
manifest later in the disease, possibly due to the depletion of CCR5
expressing T-cells \cite{Stalmeijer15032004}. For the rest of this
study we will focus only on the R5 strain of HIV-1 and consider CCR5 as
the main coreceptor. After the Env complex has bound to CCR5, it
undergoes a further conformational change that exposes gp41 which
extends and penetrates the host cell membrane \cite{PMC3405824, White2008},
bringing the cellular and viral membranes to close proximity and
allowing them to fuse. After successful viral entry, the capsid coat
dissolves and HIV-1 RNA is reverse transcribed to DNA, transported into the
nucleus, and finally integrated into the host DNA. The cell will now
produce the constituent parts needed to assemble more HIV-1 virions.

\begin{figure*}[t!]
\includegraphics[width=6.4in]{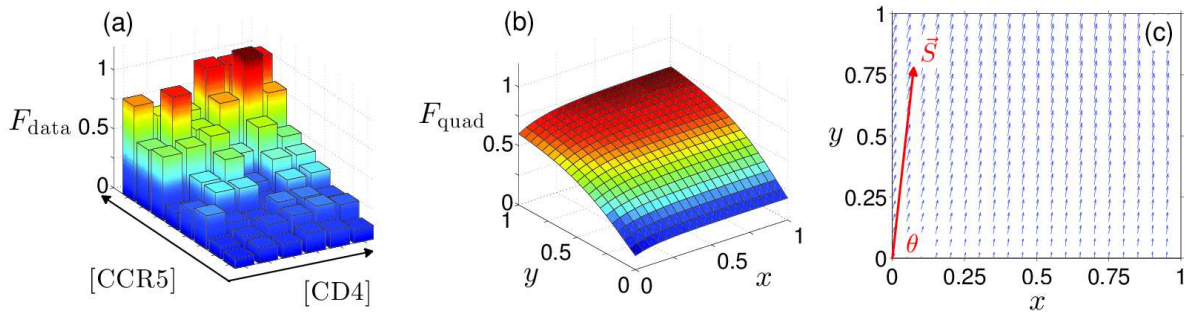}
    \caption{Quantitative analysis of infectivity measurements over a
      matrix of CD4 and CCR5 expression levels on Affinofile
      cells. The displayed data is a measurement of the infectivity of
      strain NL43(RT) \cite{Johnston}.  (a) Normalized data of relative HIV-1
      infectivity, $F_{\rm{data}}$, as a function of rescaled
      concentration of receptor CD4 and coreceptor
      CCR5. $F_{\rm{data}}$ is measured as a percentage of cells
      expressing p24 protein; an indicator of successful HIV-1
      infection. (b) Fitted quadratic function,
      $F_{\mathrm{quad}}(x,y)$, of percentage of infected cells as a
      function of rescaled CD4 and CCR5 concentrations. The average
      infectivity relative to the maximum observed is $M=52.5$. (c)
      Gradient map of the quadratic fit, $F_{\mathrm{quad}}(x,y)$,
      displays how responsiveness differs for different
      concentrations. The responsivity magnitude here is $\Delta=
      68.2$ and the responsivity angle is $\theta = 81.7^{\circ}$.}
\label{FIG1}
\end{figure*}

Since binding of surface receptor CD4 and coreceptor CCR5 are
fundamental steps in viral entry, we expect the infectivity of most
strains of HIV-1 to be particularly responsive to the cell surface
density of those receptors. This response has been investigated using
the 293-Affinofile cell line system \cite{Chikere201381, Johnston}.
%
%
%
Affinofile cells are a CD4/CCR5 dual-inducible cell line capable of
expressing independent combinations of surface expression of CD4 and
CCR5 \cite{Johnston}.  CD4 expression is induced with minocycline or
doxycycline, protein synthesis inhibitors used in antibiotics, and
CCR5 is induced with ponasterone A, an ecdysteroid activity
inducer. Once induced, Affinofile cells can be infected with
reporter-pseudotyped HIV-1 particles or live virus in a spinoculation
protocol where virions are exposed to a layer of plated cells.
Infection is then quantified through reported expression or
intracellular staining for expression of p24, the capsid protein HIV-1
uses to form a protein coat \cite{Mascola,Johnston, Webb}. By following
this protocol, Johnston \textit{et al.} \cite{Johnston} measured the
infectivity of a number of HIV-1 strains on cells that expressed a
matrix of varying levels of CD4 and CCR5. Once the levels of p24 are
measured, viral infectivity can be directly related to the associated
CD4 and CCR5 concentrations used.

In their analysis, Johnston \textit{et al.} \cite{Chikere201381} argued
that viral infectivity as functions of CD4 and CCR5 concentrations,
[CD4] and [CCR5] respectively, could be qualitatively fit to a
quadratic polynomial function 

\begin{equation}
F_{\rm{quad}}(x,y) =
a+bx+cy+dx^2+ey^2+fxy,
\label{FQUAD}
\end{equation}
representing the amount of viral particle entry measured through the
percentage of cells that are p24+. The independent variables $x$ and
$y$ are rescaled concentrations of CD4 and CCR5, respectively, defined
as:

\begin{eqnarray}
\label{XY}
x =
\frac{\log\left(\frac{[\rm{CD4}]}{[\rm{CD4}]_{\rm{min}}}\right)}{\log\left(\frac{[\rm{CD4}]_{\rm{max}}}{[\rm{CD4}]_{\rm{min}}}\right)},
\quad y =
\frac{\log\left(\frac{[\rm{CCR5}]}{[\rm{CCR5}]_{\rm{min}}}\right)}{\log\left(\frac{[\rm{CCR5}]_{\rm{max}}}{[\rm{CCR5}]_{\rm{min}}}\right)}.
\end{eqnarray}

In Eqs.~\ref{XY}, $[\rm{CD4}]_{\rm{min}}$, $[\rm{CD4}]_{\rm{max}}$ and
$[\rm{CCR5}]_{\rm{min}}$, $[\rm{CCR5}]_{\rm{max}}$ are the minimum and
maximum expression levels of receptor and coreceptor, respectively,
used in a given measurement. The rescaling in Eqs.~\ref{XY} restricts
$x$ and $y$ to be between $0$ and $1$ and is used to compare results
obtained from different experiments and/or protocols that may have
yielded different absolute ranges of [CD4] and [CCR5]. The parameters
$a$,$b$,$c$,$d$,$e$, and $f$ are estimated from fitting
$F_{\rm{quad}}$ to data.  Three metrics were derived from
$F_{\rm{quad}}(x,y)$: the mean relative infectivity $M$, and the
amplitude $\Delta$ and angle $\theta$ of the average infectivity
gradient, $\vec{S}=\int_0^1\int_0^1\vec{\nabla}F_{\rm{quad}}(x,y)\dd x
\dd y = S_x\hat{x}+S_y\hat{y}$. These were defined as:

\begin{eqnarray}
M & = &\int_0^1\int_0^1F_{\rm{quad}}(x,y)\dd x \dd y, \nonumber \\
\Delta &= &\vert\vec{S}\vert, \nonumber  \\
\theta &= &\tan^{-1}\left(\frac{S_x}{S_y}\right). 
\label{OLDMETRICS}
\end{eqnarray}
$M$ characterizes of the overall relative infectivity
of a strain of HIV-1, while $\Delta$ and $\theta$ quantify how
responsive a given strain is to receptor and coreceptor
concentrations. For example, a $\theta$ value close to $0^{\circ}$
implies $S_x\ll S_y$ and indicates infectivity that is very responsive
to CD4 expression levels, while a value close to $90^{\circ}$ implies
$S_x\gg S_y$ and indicates a high responsiveness to CCR5 levels, as
shown in Fig.~\ref{FIG1}. Johnston \textit{et al.} \cite{Chikere201381}
observed this latter pattern in several virus strains that exhibited
apparent unresponsiveness to CD4, even at extremely low CD4
concentrations, leading to the possibility of an effectively
``CD4-independent'' pathway for viral entry. A complete mechanistic
picture of the molecular processes involved however is lacking and the
current empirical evidence is insufficient to conclusively argue for a
completely CD4-independent entry pathway. Although fitting data to
$F_{\rm{quad}}(x,y)$ provides a functional relationship between
receptor and coreceptor concentrations and viral infectivity, and
descriptive parameters characterizing that relationship, the fitted
function $F_{\rm{quad}}(x,y)$ offers no mechanistic insight into the
relevant biochemical processes or their rates.  Essentially, the
parameters $M$, $\Delta$, and $\theta$ are not directly related to
mechanistic parameters involved in viral entry, especially those that
could be potentially altered by fusion inhibitors or other
therapeutics. Finally, although in previous work different viral
strains were tentatively clustered as a function of their receptor and
coreceptor usage patterns \cite{Chikere201381,Johnston}, a
mechanistic model, where kinetic rates carry a biochemical
significance, allows us to place greater confidence in the validity of
experimentally derived rates, especially if estimates from different
viral strains cluster in parameter space.

%
%
%

In this paper, we seek to quantitatively characterize HIV-1
infectivity as a function of cell surface concentrations of CD4
receptor, CCR5 coreceptor, and the associated kinetic rates.  We
propose three alternative models for receptor/coreceptor engagement
and validate them against experimental data derived from the
Affinofile cell system and derive estimates for kinetic parameters
using maximum likelihood estimation (MLE). Furthermore, we cluster the
parameter estimates from experiments derived from the same viral
strains to demonstrate confidence in our inference. Lastly, we
consider model selection criteria to compare the performance of our
proposed models and assess their utility in modeling HIV-1 infection.

\section{Theory}

\subsection{Sequential Model}
\begin{figure*}[ht]
\begin{center}
\includegraphics[width=4.5in]{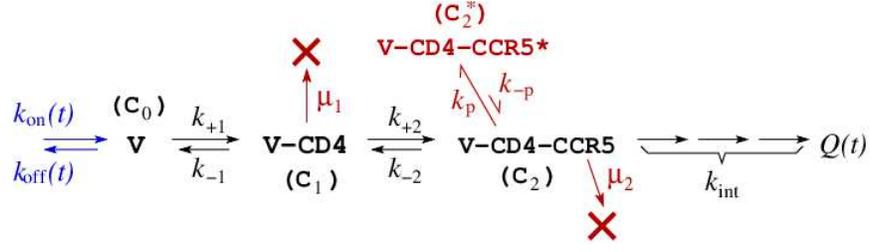}
\caption{Sequential kinetic model of viral entry. HIV-1 viruses
  that are associated with the host membrane, V, are adsorbed with
  rate $k_{\rm{on}}$ and dissociate with rate $k_{\rm{off}}$. They can
  then bind to CD4 receptors to become V-CD4 with rate $k_{+1}$ from
  which they can unbind with rate $k_{-1}$ or degrade with rate
  $\mu_1$. The V-CD4 complex can bind to coreceptor CCR5 to become
  V-CD4-CCR5 with rate $k_{+2}$, reverse the process with rate
  $k_{-2}$, degrade with rate $\mu_2$, or carry on to full cell
  membrane fusion and integration with rate $k_{\rm{int}}$. We include
  the peptide-bound state V-CD4-CCR5* to factor in fusion inhibition
  which would sequester the CD4 and CCR5 bound viruses with rate
  $k_{\rm{p}}$ and degrade with rate $\mu_{\rm p}$.}
\label{SERIESFIG}
\end{center}
\end{figure*}

The simplest model of HIV-1 viral entry is based on the assumption that
binding of the viral Env protein complex to cell surface receptor CD4
is a necessary precursor step for viral
entry \cite{PMC3405824,Boulant,Platt1}. This binding causes Env to
undergo a conformational change that allows further binding to the
CCR5 coreceptor, initiating the fusion event. We display this
``sequential'' binding assumption with rate parameters and pathways in
Fig.~\ref{SERIESFIG}. We denote the concentration [V] (number per host cell area)
of membrane-associated HIV-1 virion particles which are not bound to any
receptors by $c_0(t)$, the concentration [V-CD4] of CD4 receptor-bound
HIV-1 by $c_1(t)$, and the concentration [V-CD4-CCR5] of HIV-1 bound to
both CD4 and CCR5 by $c_2(t)$, at a given time $t$. In addition, we
include a potential fusion inhibitor and denote by $c_{2}^{*}(t)$ the
concentration [V-CD4-CCR5*] of HIV-1 that is bound to CD4 and CCR5 and
to an external peptide that impedes fusion, effectively sequestering
the cell from further progressing towards infection.
Although there are many intermediate steps during membrane fusion and
inside the cytoplasm that ultimately result in viral DNA integration,
we subsume these processes into a single step that follows the
assembly of the V-CD4-CCR5 complex in the rate parameter
$k_{\rm{int}}$. We also assume the adsorption rate, $k_{\rm{on}}(t)$,
of free virus onto the cellular surface is time dependent since cell
adhesion is high during spinoculation when the HIV-1 viruses are driven
close to the cell membrane \cite{Chou3}. After spinoculation, the
culture medium is replaced to wash away free virus particles. Therefore, 
for times $t>0$ adsorption of new HIV-1 to the membrane is precluded
and we set $k_{\rm{on}}(t> 0)=0$. Finally, while $k_{\rm{off}}$
describes the rate of HIV-1 desorption from the cell membrane, $\mu_1$
and $\mu_2$ describe the rate of CD4 or CCR5-bound virus elimination
via capsid protein coat degradation, endocytosis, or other abortive
events. Using these assumptions, we can mathematically describe the
sequence of events leading to infection for $t> 0$ as follows:
\begin{eqnarray}
\frac{\dd c_0(t)}{\dd t} &=& -k_{+1}c_0 - k_{\rm{off}}c_0 + k_{-1}c_1,
\nonumber \\ 
\frac{\dd c_1(t)}{\dd t} &=& k_{+1}c_0 - k_{-1}c_1 - \mu_1c_1 -
k_{+2}c_1 + k_{-2}c_2, \\
\frac{\dd c_2(t)}{\dd t} &=& k_{+2}c_1 - k_{-2}c_2 - k_{\rm{p}}c_2 -
k_{\rm{int}}c_2 - \mu_2 c_2 + k_{-\rm{p}} c_{2}^{*}, \nonumber \\ 
\frac{\dd c_{2}^{*}(t)}{\dd t} &=& k_{\rm{p}} c_2 - k_{-\rm{p}}c_{2}^{*} -
\mu_{\rm p} c_{2}^{*} \nonumber.
\label{SERIESEQN}
\end{eqnarray}
The above equations represent the concentration flow in and out of the
four states the virus can inhabit, V, V-CD4, V-CD4-CCR5, and
V-CD4-CCR5*, respectively, as detailed in Fig.~\ref{SERIESFIG}. For
example, the three terms in the first equation, from left to right,
describe HIV-1 binding to CD4, HIV-1 desorbing from the cell membrane, and
CD4-bound HIV-1 dissociating from CD4 while maintaining cell adhesion.
We adopt the simplest assumption that the overall receptor
and coreceptor binding rates $k_{+1}$ and $k_{+2}$ are increasing functions 
of the surface expression of CD4 and CCR5, respectively, and 
define

\begin{equation}
k_{+1} = k_{+1}^0[\mathrm{CD4}]^{\beta_1}\quad \mbox{and}\quad 
k_{+2} = k_{+2}^0[\mathrm{CCR5}]^{\beta_2}, 
\label{KK0}
\end{equation}
where $k_{+1}^0$ and $k_{+2}^0$ are the intrinsic binding rates
between Env and the respective receptors CD4 and CCR5. The
stoichiometry, $\beta_1$ and $\beta_2$, represent, in the infinitely
cooperative binding limit, the number of receptors and coreceptors
that must bind before fusion can be
triggered \cite{Weiss01091997,Kuhmann}.  For example, Env is known to
form a trimer of gp120/gp41 complexes, with each subunit containing a
CD4 binding domain \cite{PMC3405824,Zolla-Pazner,Platt1}. If typically
two or more of these binding domains are required to be bound to CD4
receptors before the appropriate conformational changes occur, we
expect the associated Hill coefficient $\beta_1>1$.
\begin{figure}[ht]
\includegraphics[width=3.5in]{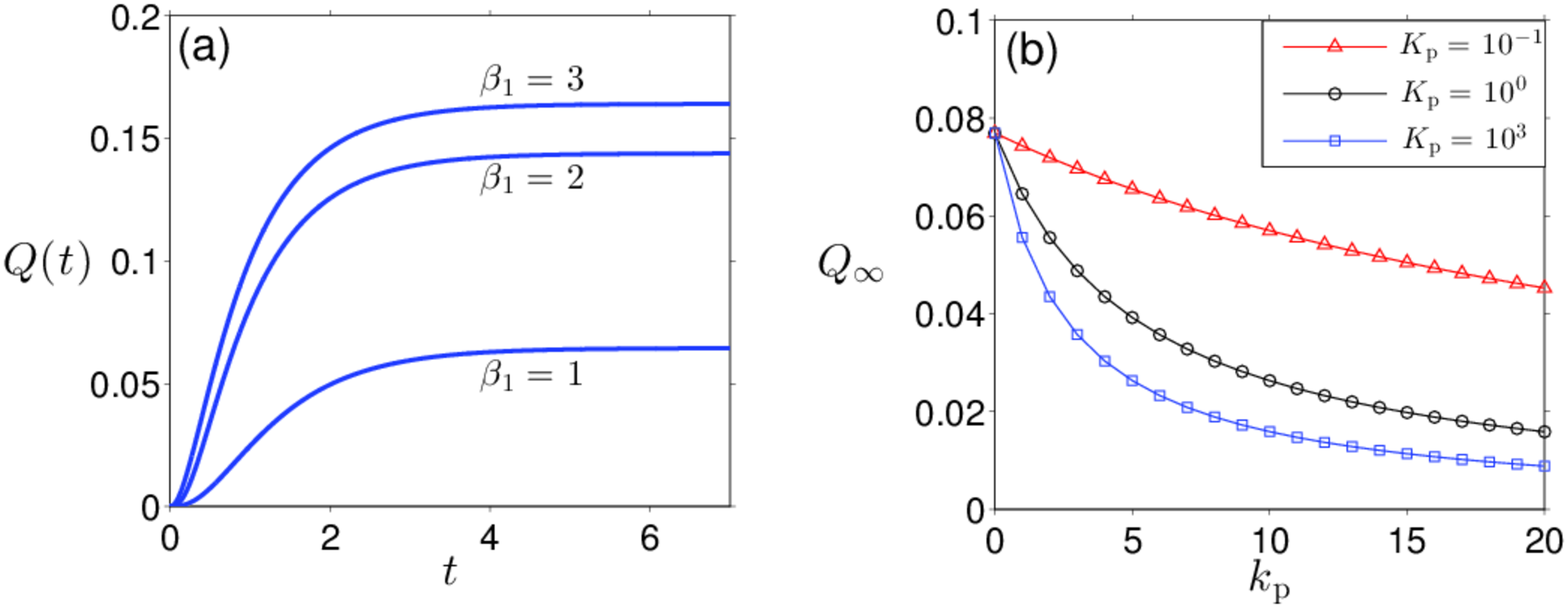}
    \caption{(a) Infectivity $Q(t)$ calculated from Eq.~\ref{Q} using
      different values of $\beta_1$ and setting all other rates to
      $1\rm{s}^{-1}$. Since the Env trimer has three possible binding
      domains for CD4, infectivity will increase if the complex
      exhibits infinite binding cooperativity across two or three of
      the subunits.  (b) Effectiveness of fusion inhibitor on total
      viral infectivity $Q_{\infty}$ defined in Eq.~\ref{QINFTY} as a
      function of varying peptide binding rate $k_{\rm{p}}$. We plot
      the resulting curves for different values of $K_{\rm{p}} =
      \mu_{\rm{p}}/k_{-\rm{p}}$ while setting all other rate
      parameters to $1\rm{s}^{-1}$ and $\beta_1=\beta_2=1$. Since
      $k_{\rm{p}}$ scales with the concentration of fusion inhibitor
      peptide, increasing the latter inhibits HIV-1 infectivity more
      effectively at lower peptide concentrations than at higher
      ones.}
\label{QT}
\end{figure}

Within our mathematical model, we represent HIV-1 infectivity by
\begin{equation}
Q(t) = \int_0^t k_{\rm{int}}c_2(\tau)\dd\tau,
\label{Q}
\end{equation}
the fraction of initially-adsorbed virus particles that that have
undergone fusion by time $t$. $Q(t)$ represents the cumulative number
of successful fusion events from state V-CD4-CCR5. Once the relevant
rates are determined, given an initial concentration of adsorbed
virus, $c_0(0)\equiv V_{0}$, and assuming no other bound complexes so
that $c_1(0)=c_2(0)=c_3(0)=0$, we can derive $Q(t)$ from
Eqs.~\ref{SERIESEQN} and compare analytical results with the
experimental measurements of viral infection on Affinofile cells
\cite{Webb,Chikere201381}. The qualitative behavior of $Q(t)$ for
various $\beta_{1}$ is shown in Fig.~\ref{QT}(a) assuming a
hypothetical case where all rates are set to 1s$^{-1}$.  The plot
shows an initial steep increase of viral infectivity immediately after
$t=0$ as viruses progress toward receptor and coreceptor binding and
fusion. Eventually the initial concentration of HIV-1, $V_0$, is
depleted, at which point $Q(t)$ flattens out. We expect larger values
of $\beta_1$ to yield larger values of $Q(t)$ since they allow for
stronger binding, while the $\beta_1$-independent dissociation and
degradation rates stay the same. Since viral infectivity is measured
after a sufficiently long exposure time we focus on the long time
value $Q_{\infty} \equiv \lim_{t \to \infty}Q(t)$. Upon solving
Eqs.~\ref{SERIESEQN} and Eq.~\ref{Q}, we find

\begin{widetext}
\begin{eqnarray}
Q_{\infty} &=& \displaystyle 
\frac{V_0}{\left(\frac{\omega-k_{-2}}{k_{\rm{int}}}\right) +
  \left(\frac{k_{\rm{off}}\left(\omega -
    k_{-2}\right)}{k_{+1}k_{\rm{int}}}\right) + \left(\frac{\omega
    \mu_1}{k_{+2}k_{\rm{int}}}\right) + \left(\frac{\omega
    k_{\rm{off}}\left(k_{-1}+\mu_1\right)}{k_{+1}k_{+2}k_{\rm{int}}}\right)},
\label{QINFTY}
\end{eqnarray}
\end{widetext}
where
\begin{equation}
\omega = k_{-2} + k_{\rm{int}} + \left(\frac{\mu_{\rm{p}}}{k_{\rm{-p}}+\mu_{\rm{p}}}\right)k_{\rm{p}} 
+ \mu_2.
\end{equation}
The quantity $\omega$ can be interpreted as the total flow of virus
out of the state V-CD4-CCR5 as the individual terms describe, from
left to right, the dissociation of CCR5, viral membrane fusion, fusion
inhibiting peptide binding, and V-CD4-CCR5 complex degradation. Note
that, through the rate parameters $k_{+1}$ and $k_{+2}$
(Eqs.~\ref{KK0}), $Q_\infty$ is a function of the concentrations [CD4]
and [CCR5].

As evident from Eq.~\ref{QINFTY}, the total infectivity depends on
rates of known kinetic processes and initial viral
concentrations. Measuring the effects of each of these parameters on
$Q_\infty$ is particularly useful in the study and development of drug
therapies. For example, if we make the reasonable assumption that the
fusion inhibitor peptide binding rate $k_{\rm{p}}$ is proportional to
the concentration of the peptide in the extracellular environment, we
may vary $k_{\rm{p}}$ while keeping all other parameters fixed and
observe how infectivity changes, as depicted by the dose-response curves
in Fig.~\ref{QT}(b). Here, increasing $k_{\rm{p}}$ leads to a more
pronounced decrease in infectivity $Q_\infty$ at low $k_{\rm{p}}$,
while for high values of $k_{\rm{p}}$, changes in $Q_\infty$ are less
significant.  Thus, our analyses of the model can be used to guide,
based on mechanistic principles, the development and administration of
entry inhibitor therapeutics.

\section{Results \& Discussion}

In order to compare experiments from a number of different HIV-1
strains, cell lines, and laboratory conditions, rescaled expression
levels of CD4 and CCR5, $x$ and $y$, respectively, are used.  Solving
Eqs.~\ref{XY} for [CD4] and [CCR5], we can express the rates $k_{+1}$
and $k_{+2}$ as functions of $x$ and $y$.  We find
$k_{+1}=k_{+1}^0[\mathrm{CD4}]^{\beta_{1}}
=k_{+1}^0[\mathrm{CD4}]_{\mathrm{min}}^{\beta_{1}}
\left(\frac{[\mathrm{CD4}]_{\mathrm{max}}}{[\mathrm{CD4}]_{\mathrm{min}}}\right)^{\beta_{1}x}$
and
$k_{+2}=k_{+2}^{0}[\mathrm{CCR5}]^{\beta_{2}}=
k_{+2}^0[\mathrm{CCR5}]_{\mathrm{min}}^{\beta_{2}}
\left(\frac{[\mathrm{CCR5}]_{\mathrm{max}}}{[\mathrm{CCR5}]_{\mathrm{min}}}\right)^{\beta_{2}y}$.
%
%
Furthermore, as $V_0$ in Eq.~\ref{QINFTY} has units of concentration,
so does $Q_{\infty}$. To nondimensionalize the expression in
Eq.~\ref{QINFTY} and allow for direct comparisons of results across
different experiments, we can normalize $Q_{\infty}$ by a reference
virus concentration. Experimentalists commonly use the raw infectivity
value corresponding to the highest concentrations of CD4 and CCR5 as
the reference concentration \cite{Johnston,Roche2011}. If we
define $Q_{\rm{max}}$ as the experimental infectivity at
$[\rm{CD4}]_{\rm{max}}$ and $[\rm{CCR5}]_{\rm{max}}$, the normalized
infectivity becomes

\begin{eqnarray}
F_{\infty}(x,y\vert \boldsymbol\xi)& \equiv & \frac{Q_{\infty}}{Q_{\rm{max}}} \nonumber \\
\: & = & \frac{D}{1+AX^{\beta_{1}x}+BY^{\beta_{2}y}+ABC X^{\beta_{1}x}Y^{\beta_{2}y}},
\label{QINFTY2}
\end{eqnarray}
where $X=\frac{[\rm{CD4}]_{\rm{min}}}{[\rm{CD4}]_{\rm{max}}}$ and
$Y=\frac{[\rm{CCR5}]_{\rm{min}}}{[\rm{CCR5}]_{\rm{max}}}$ are
experiment-dependent constants and $\boldsymbol\xi \equiv
\{A,B,C,D,\beta_1,\beta_2\}$ is a six-dimensional vector of
parameters. The dimensionless combination of parameters
$A$, $B$, $C$ and $D$ are defined by

\begin{align}
A & = \left(\frac{1}{[\rm{CD4}]_{\rm{min}}}\right)^{\beta_{1}}\left(\frac{k_{\rm{off}}}{k_{+1}^0}\right), \nonumber \\
B & = \left(\frac{1}{[\rm{CCR5}]_{\rm{min}}}\right)^{\beta_{2}}
\frac{\mu_1\omega}{k_{+2}^0\left(\omega-k_{-2}\right)}, \label{ABCD} \\
C & =  \left(1+\frac{k_{-1}}{\mu_1}\right), \nonumber \\
D & =  {k_{\rm int}V_{0}\over Q_{\rm{max}}(\omega - k_{-2})}. \nonumber
\end{align}

\begin{figure*}[ht]
\includegraphics[width=6.5in]{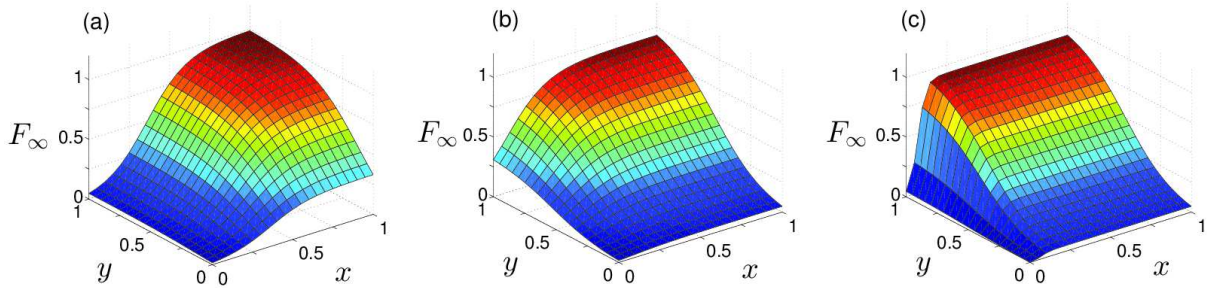}
    \caption{$F_{\infty}(x,y\vert \boldsymbol\xi)$ for different sets
      of $\boldsymbol\xi$. (a) For $(A, B, C, D, \beta_{1}, \beta_{2})
      = (20, 2, 1, 1, 2,2)$, the infectivity is most sensitive to $x$,
      or [CD4]. (b) For $(A, B, C, D, \beta_{1}, \beta_{2}) = (2, 20,
      1, 1, 2,2)$, the function $F_{\infty}$ varies more along the $y$
      direction and infectivity is more sensitive to [CCR5].  (c) $(A,
      B, C, D, \beta_{1}, \beta_{2}) = (20, 20, 1, 1, 10,2)$. Note
      that $A$ and $B$ predominately control the direction (CD4 or
      CCR5) of sensitivity, and $\beta_1$, $\beta_2$ control the steepness
      of the infectivity surface.}
\label{QINFTY_SURFACES}
\end{figure*}

For illustration, in Fig.~\ref{QINFTY_SURFACES} we plot
$F_{\infty}(x,y\vert \boldsymbol\xi)$ for three different sets of
$\boldsymbol\xi$. In Fig.~\ref{QINFTY_SURFACES}(a), we assume a small
value of the dimensionless parameter $B$, while in
Fig.~\ref{QINFTY_SURFACES}(b) we assume a small dimensionless
parameter $A$. In Fig.~\ref{QINFTY_SURFACES}(c), we set $A=B$ but we
assume different stoichiometries $\beta_{1}=5$ and $\beta_{2}=1$. All
values of the chosen parameters $\boldsymbol\xi$ are presented in the
figure caption. The different sets of parameters yield model
infectivity functions $F_{\infty}(x,y\vert \boldsymbol\xi)$ with
gradients along various directions in the $(x,y)$ plane, indicating
greater sensitivity to [CD4] or [CCR5]. Furthermore, higher values of
stoichiometry $\beta_{1}, \beta_{2}$ amplify the sensitivity over a
range of concentration values. Therefore, $F_{\infty}(x,y\vert
\boldsymbol\xi)$ may effectively represent different viral strains
with different CD4/CCR5 usage patterns distinguished by values of
$\boldsymbol\xi$.

Before fitting to data to find the MLE of $\boldsymbol\xi$, we note
that even though $Q_{\rm{max}}$ is a known constant from the raw
infectivity data, some previous data sets do not report its
value \cite{Johnston,Roche2011}. Therefore this normalization factor is
subsumed into the inference of $D$ but still can be inferred provided
the other parameters forming $D$ can be independently determined.

If we wish to fit raw unnormalized infectivities, we can also use the
form for $F_{\infty}(x,y\vert \boldsymbol\xi)$ directly, but redefine
the prefactor $D = {k_{\rm int}V_{0}\over (\omega - k_{-2})}$ as
having the same units as the experimental output (such as number of
cells, fluorescence, etc.). In this case, the parameter $D$ must be
rescaled by an experimental factor with units of the experimental
output multipled by an area. In either case, we consider the amplitude
$D$ as a free parameter to be inferred from data fitting.  This
parameter incorporates, and is confounded by, the initial intensity of
exposure, as well as many of the sample-to-sample experimental
variability arising from instrument error, host cell number, and area
within each sample well.  Therefore, we do not expect the effective
values of $D$ to systematically represent intrinsic kinetic
rates. However, we do expect the remaining parameters
$A,B,C,\beta_{1}$, and $\beta_{2}$ to influence the shape of
$F_{\infty}(x,y \vert \boldsymbol\xi)$ and the receptor/coreceptor
usage patterns.  Note that in addition to the intrinsic rate
parameters, $A$ and $B$ also depend on [CD4]$_{\rm min}$ and
[CCR5]$_{\rm min}$ which can vary from from one measurement to the
next.  Therefore, direct comparisons of $A$ and $B$ can be made only
across measurements that use the same $[\mathrm{CD4}]_{\mathrm{min}}$
and $[\mathrm{CCR5}]_{\mathrm{min}}$.

We now fit $F_{\infty}(x,y\vert \boldsymbol\xi)$ to
previously-obtained normalized infectivity by finding the maximum
likelihood values for $\hat{\boldsymbol\xi}=
(\hat{A},\hat{B},\hat{C},\hat{D},\hat{\beta}_1,\hat{\beta}_2)$. The
appropriate likelihood function is based on the assumption that
chemical kinetic rates are typically products of positive random
variables representing chemical concentrations and other rates.  This,
and the need to restrict the infectivity signal to strictly positive
values implies that a log-normal distribution for values of
$F_{\infty}$ is reasonable. The maximum likelihood estimation for
$\boldsymbol\xi$ then equivalent to minimizing the objective function

\begin{equation}
\Phi(\boldsymbol\xi) = \sum_{i,j}\left(\log F_{\infty}(x_i, y_j\vert
  \boldsymbol\xi) - \log F_{\rm{data}}(x_i,
  y_j) \right)^2,
\label{OBJECTIVE}
\end{equation}

\begin{figure*}[ht]
\includegraphics[width=6.5in]{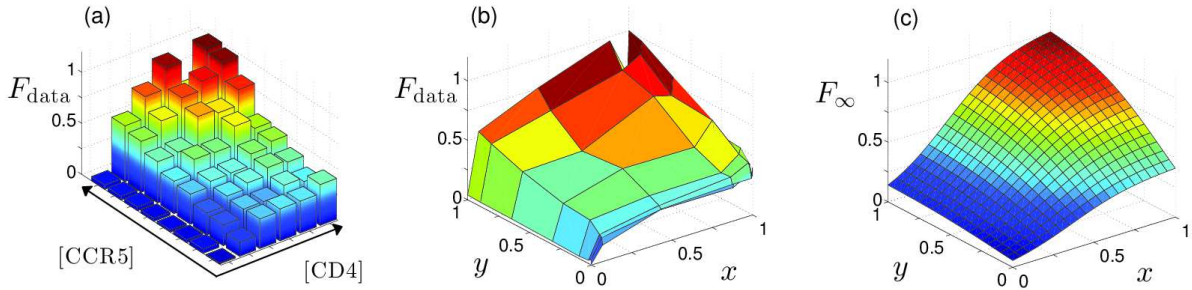}
    \caption{HIV-1 infectivity as a function of CD4 and CCR5
      concentrations. The data displayed is from strain B5(YA)
      \cite{Johnston}. (a) Normalized unscaled infectivity data
      measured after a sufficiently long exposure time. (b) Normalized
      and rescaled infectivity data in terms of $(x,y)$.  (c) Fitted
      plot of scaled normalized infectivity
      $F_\infty(x,y\vert\hat{\boldsymbol\xi})$ from Eq.~\ref{QINFTY}
      assuming $C=1$ and $\beta_1=\beta_2=1$. The MLE of the remaining
      parameters are $\hat{A}=6.39$, $\hat{B}=1.56$, and
      $\hat{D}=1.09$.}
\label{FIXEDBETAFIT}
\end{figure*}

\noindent
where $F_{\rm{data}}(x_i, y_j)$ are measured values of normalized
infectivity at rescaled concentrations $x_i$ and $y_j$ of CD4 and CCR5
used in the experiments. As a first approximation, we assume $\beta_1
= \beta_2 = 1$ and $C\approx 1$. This last approximation is valid when
the CD4 dissociation rate $k_{-1}$ is much smaller than the CD4-bound
degradation rate $\mu_{1}$, a chemically reasonable assumption.  In
our subsequent fits using all parameters
$(\hat{A},\hat{B},\hat{C},\hat{D},\hat{\beta}_1,\hat{\beta}_2)$, the
best-fit value of $\hat{C}$ is indeed near one. Therefore by
henceforth setting $\beta_1 = \beta_2 = 1$ and $C=1$, we reduce the
number of parameters to be estimated from six to three. The results
obtained by fitting the data from several experimental measurements of
HIV-1 strains can be found in Table 1.

\begin{figure}[ht]
\includegraphics[width=3.5in]{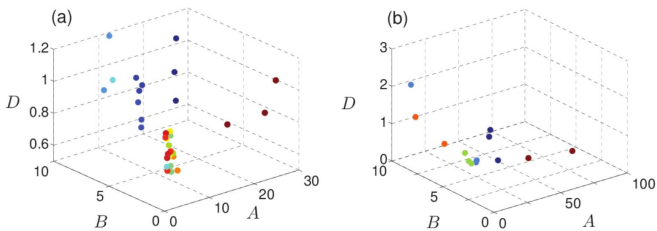}
\caption{MLE points in $A-B-D$ parameter space in the context of the
  sequential kinetic model assuming $C=1$. (a) Fixed
  $\beta_1=\beta_2=1$. Each point represents an experiment for which
  parameters were estimated. Measurements were derived from published
  data \cite{Johnston,Roche2011,Salimi2013} and points corresponding to
  replicate measurements on the same viral strain are shown in the
  same color. There is large variation in the inferred parameters
  between viral strains, but parameters inferred from replicate
  measurements on the same strain cluster relatively closely in
  $(A,B)$ parameter space. As expected, there is high measurement
  variability in $D$. (b) Allowing $\beta_1$ and $\beta_2$ to be free
  parameters to be estimated from fitting. Here variability in the
  $A$, $B$, and $D$ parameters increase as they become less sensitive
  to the functional form presented in Eq.~\ref{QINFTY2} to compensate
  for the high sensitivity of the stoichiometric parameters $\beta_1$
  and $\beta_2$. Here, we include only representative data points that
  clustered, confirming the large variability in parameter estimates.}
\label{CLUSTERFIG}

\end{figure}

In Fig.~\ref{FIXEDBETAFIT} we plot the fitted curve of
$F_\infty(x,y\vert\hat{\boldsymbol\xi})$ using the estimated
parameters from viral strain B5(YA) shown in the fourth line of Table
1, which qualitatively shows good agreement with the corresponding
experimental data. Upon repeating the same analysis on 32 experimental
data sets selected from previous
publications \cite{Johnston,Chikere201381}, we can associate each set
of infectivities with fitted $\hat{A}$-$\hat{B}$-$\hat{D}$ values. As
expected, infectivities of the same strain of HIV-1 cluster together
in $A$-$B$ parameter space as shown in Fig.~\ref{CLUSTERFIG}. To
indicate the goodness of fit to our model, we also calculate and
display in Table 1 the coefficient of determination 

\begin{equation}
R^{2} \equiv 1 - {\sum_{i,j}\left(\log F_{\infty}(x_{i}, y_{j}\vert \hat{\boldsymbol\xi}) 
- \log F_{\rm data}(x_{i}, y_{j})\right)^{2}\over
\sum_{i,j}\left(\log \bar{F}_{\rm data} - \log F_{\rm data}(x_{i}, y_{j})\right)^{2}},
\label{R2}
\end{equation}
where $\bar{F}_{\rm data}$ is the measured normalized infectivity
averaged over all [CD4] and [CCR5] concentration combinations.  The
low $R^2$ values suggest that our initial assumption of
$\beta_1=\beta_2=1$ should be relaxed to obtain better fits.

\begin{table}[h!]
\caption{Fitted parameters of sequential kinetic model
  assuming $C=1$ and $\beta_1=\beta_2=1$ for six different experiments
  representing triplicate measurements of two HIV-1 strains. The
  parameters are in close agreement within each viral strain. $R^2$
  values are shown with low values implying the $\beta_1=\beta_2=1$
  assumption is overly constraining.}
\label{tab:title1}
\begin{tabular}{l|*{6}{c}r}
\hline
Strain-Experiment              & $\hat{A}$ & $\hat{B}$ & $\hat{D}$ & $R^2$\\
\hline
B5 (YA)-Rep1 & 6.39 & 1.56 & 1.09 & 0.64\\
B5 (YA)-Rep2            & 6.03 & 1.53 & 1.27 & 0.64\\
B5 (YA)-Rep3           & 5.91 & 1.34 & 1.49 & 0.66\\
B5 (RT)-Rep1     & 3.77 & 3.80 & 1.1 & 0.60\\
B5 (RT)-Rep2     & 4.18 & 4.25 & 1.17 & 0.68 \\
B5 (RT)-Rep3     & 4.58 & 3.87 & 1.13 & 0.64 \\
\hline 
\end{tabular}
\end{table}

\begin{figure*}[ht]
\includegraphics[width=6.5in]{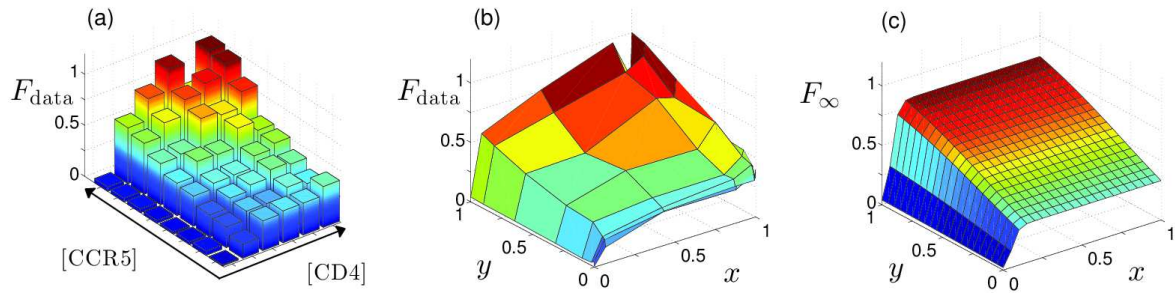}
    \caption{Normalized HIV-1 infectivity as a function of CD4 and CCR5
      levels. The data displayed is from strain B5(YA)
      \cite{Johnston}. (a) and (b) are identical to the first two
      panels in Fig.~\ref{FIXEDBETAFIT}. (c) Fitted plot of scaled
      normalized infectivity $F_\infty(x,y\vert\hat{\boldsymbol\xi})$
      with $C=1$ but free $\beta_1,\beta_2$. The MLE of the parameters
      are $\hat{A}=17$, $\hat{B}=1.84$, $\hat{D}=0.96$,
      $\hat{\beta}_{1}=11.7$, and $\hat{\beta}_{2}=0.8$.}
\label{FREEBETAFIT}
\end{figure*}

We thus consider the role of stoichiometry of receptor and coreceptor
binding by reintroducing $\beta_1$ and $\beta_2$ as free parameters
and perform maximum likelihood fitting using five parameters
$\boldsymbol\xi=\left(A,B,D,\beta_1,\beta_2\right)$, again using the
approximation $C=1$. The fitted function is shown in
Fig.~\ref{FREEBETAFIT} which is qualitatively different from
Fig.~\ref{FIXEDBETAFIT}.

The new estimated parameters using the same data as before are
displayed in Table 2. Here, the residual values $R^2$ are higher,
indicating a much better fit of the data when $\beta_{1},\beta_{2}$
are adjustable. The kinetic model $F_\infty(x,y\vert\boldsymbol\xi)$
in Eq.~\ref{QINFTY} consistently outperforms the quadratic model
$F_{\rm{quad}}(x,y)$ introduced earlier.

\begin{figure}[t!]
\includegraphics[width=3.6in]{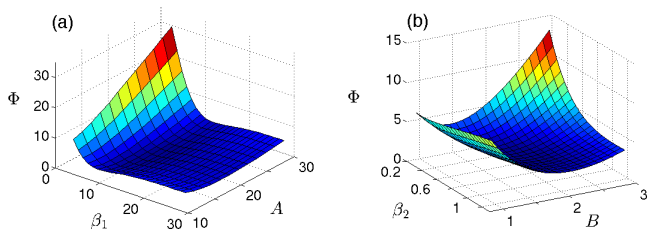}
    \caption{Projections of $\Phi(\boldsymbol\xi)$ about the minimum
      at $(\hat{A}, \hat{B}, \hat{C}, \hat{D}, \hat{\beta}_{1},
      \hat{\beta}_{2}) = (17,1.84,1,0.96,11.7,0.8)$. (a) Projection on
      $(A,\beta_{1})$ space. (b) Projection on $(B,\beta_{2})$ space.}
\label{PHIFIG}
\end{figure}

\begin{table*}[t!]
\caption{Fitted parameters of the sequential
    kinetic model assuming $C=1$ for six sets of measurements, three
    replicates of each of two different strains of HIV-1. The MLE
    parameter values are in close agreement within each viral
    strain. $R^2$ values are calculated for both the arbitrary
    quadratic model introduced earlier and the sequential kinetic
    model (Fig.~\ref{SERIESFIG}). The mechanism-based kinetic model
    consistently outperforms the quadratic model.} \label{tab:title2}
\begin{tabular}{l|*{7}{c}r}
\hline
Strain-Experiment & $\hat{A}$ & $\hat{B}$ & $\hat{D}$ & $\hat{\beta}_1$ & $\hat{\beta}_2$ &$R^2$: $F_{\mathrm{quad}}$& $R^2$:$F_\infty$\\
\hline
B5 (YA)-Rep1 & 17.04 & 1.84 & 0.96 & 11.7 & 0.79 & 0.95 & 0.96 \\
B5 (YA)-Rep2            & 16.40 & 2.62 & 1.49 & 11.6 & 0.57 & 0.94 & 0.98 \\
B5 (YA)-Rep3           & 14.85 & 2.30 & 1.73 & 10.8 & 0.55 & 0.80 & 0.97 \\
B5 (RT)-Rep1     & 10.67 & 3.15 & 0.79 & 13.9 & 1.29 & 0.79 & 0.90 \\
B5 (RT)-Rep2     & 9.90 & 9.42 & 2.01 & 10.7 & 0.68 & 0.77 & 0.92 \\
B5 (RT)-Rep3     & 12.32 & 3.30 & 0.81 & 12.1 & 1.23 & 0.81 & 0.92 \\
\hline 
\end{tabular}
\end{table*}
The value of $\hat{\beta}_2$ is consistently close to $1$ which
indicates that coreceptor binding only involves a single CCR5 for
fusion to be initiated. The value of $\hat{\beta}_1$, on the other
hand, is much larger.  If interpreted as a binding stoichiometry, this
would indicate that multiple gp120/gp41 complexes of the Env trimer
must bind to separate CD4 receptors before conformational changes can
take place. Our results, however, show $\beta_1>3$, which indicates
that each individual gp120/gp41 complex binds to multiple CD4
receptors. Alternatively, the large exponent $\beta_1$ can describe
high allosteric cooperativity of Env or the dynamics of multiple Env
each binding to CD4, increasing the effective rate of CCR5 binding. To
quantify our confidence in this result, we take a closer look at the
objective function $\Phi(\boldsymbol\xi)$ in Eq.~\ref{OBJECTIVE} and
evaluate it for $\boldsymbol\xi$ values close to
$\hat{\boldsymbol\xi}$, the optimal value of $\boldsymbol\xi$ that
minimizes $\Phi(\boldsymbol\xi)$ constrained by the data. By varying
one of the parameters at this minimum while keeping all others fixed,
we can measure the rate of change in $\Phi(\boldsymbol\xi)$ with
respect to that parameter. In particular, we can determine the
sensitivity of the model with respect to a given parameter by
evaluating the curvature of $\Phi(\boldsymbol\xi)$, a measure of fit
error, along the direction in which that parameter changes, as shown
in Fig.~\ref{PHIFIG}. 

To compare the performance of the quadratic model described by
$F_{\rm{quad}}(x,y)$ with that of our kinetic model described by
$F_{\infty}(x,y \vert\hat{\boldsymbol\xi})$ both with
$\beta_1=\beta_2=1$ and as free parameters, we calculate the Akaike
Information Criterion (AIC) score

\begin{eqnarray}
\mathrm{AIC} & = & 2n + \sum_{i,j}\log\left(2\pi F_{\rm{data}}^2(x_i,
y_j)\right) \\
\: \hspace{1cm} & + &\sum_{i,j}\left(\log F_{\infty}(x_i, y_j\vert
\hat{\boldsymbol\xi}) - \log F_{\rm{data}}(x_i,
y_j) \right)^2,
\label{AIC}
\end{eqnarray}


\noindent a standard statistical measure for model comparison and
selection. In Eq.~\ref{AIC}, $n$ is the number of inferred parameters
in the model and the last two terms are derived from the
log-likelihood function of the infectivity distribution. The AIC score
penalizes models with large errors in the prediction of each data
point and with too many fitted parameters, so a low AIC score is
ideal. We observe that the kinetic model with $\beta_1$ and $\beta_2$
as free parameters once again outperforms both the models with fixed
$\beta_{1} = \beta_{2} = 1$ and the quadratic model, further
validating our mechanistically derived model. This implies that the
data provides some confidence in a higher stoichiometry $\beta_{1} >
1$.

\begin{table}[t!]
\caption{AIC scores, defined in Eq.~\ref{AIC}, for the
  arbitrary quadratic model, our sequential kinetic model assuming $C=1$
  and $\beta_1=\beta_2=1$, and the same model with now $\beta_1$, and
  $\beta_2$ as free parameters to be estimated. The latter model has
  shown to consistently outperforms both the fixed $\beta_{1},
  \beta_{2}$ model and the quadratic model introduced
  earlier.} \label{tab:title3}
\begin{tabular}{l|*{6}{c}r}
\hline
Strain-              &AIC:$F_{\mathrm{quad}}$ & AIC:$F_\infty$ & AIC:$F_\infty$   \\
Experiment & & ($\beta_1=\beta_2=1$) & (free $\beta_1$,$\beta_2$)\\
\hline
B5 (YA)-Rep1 & -19.9 & -7.30 & -22.4 \\
B5 (YA)-Rep2            & -1.80 & 9.20 & -6.0 \\
B5 (YA)-Rep3           & 26.0 & 28.2 & 15.6 \\
B5 (RT)-Rep1     & -31.4 & -25.7 & -39.4 \\
B5 (RT)-Rep2     & -35.5 & -36.0 & -45.2 \\
B5 (RT)-Rep3     & -38.0 & -32.9 & -46.1 \\
\hline 
\end{tabular}
\end{table}

Once accurate estimates of the model parameters $A$, $B$, $D$,
$\beta_1$, and $\beta_2$ are obtained from minimizing
$\Phi(x,y\vert\boldsymbol\xi)$, we can derive constraints on the
physical rate parameters through Eqs.~\ref{ABCD}.  Although there are
more kinetic parameters to solve for than available estimates, we can
use known rate values obtained from past ligand binding
assays \cite{jp1080725,Seisenberger30112001}. For example, Chang et
al. \cite{PMC1287567} set the Env-CD4-CCR5 dissociation rate at $k_{-2}
\approx 1.7\rm{s}^{-1}$. Furthermore, if we assume all degradation
rates are equal, we can follow Seisenburger et
al. \cite{Seisenberger30112001} and use $\mu_1=\mu_2\approx
15\rm{s}^{-1}$ as general viral degradation rates.  Furthermore,
through GFP genetic marking and flow cytometry, assays can be designed
to potentially measure the nonspecifically absorbed virus dissociation
rate $k_{\rm{off}}$ \cite{Dale2011}.


The data shown in Fig.~\ref{FIG1} reveals that the entry of the
associated viral strain is very insensitive to CD4 levels for the
values explored.  Using the previous metrics defined in
Eqs.~\ref{OLDMETRICS}, the sensitivity vector $\vec{S}$ points almost
entirely in the $y$ (CCR5) direction.  Previous work has also
suggested the existence of ``CD4-independent'' strains that infect at
extremely low levels of CD4 \cite{DOMS1999,CHENGMAYER}. However, our
sequential model requires binding of CD4 before fusion can occur. In
the next section, we explore an alternative and more general ``parallel''
pathway model that may better fit observed data such as that
illustrated in Fig.~\ref{FIG1}.


\subsection{Parallel Model}
\begin{figure}[ht]
\begin{center}
\includegraphics[width=3.0in]{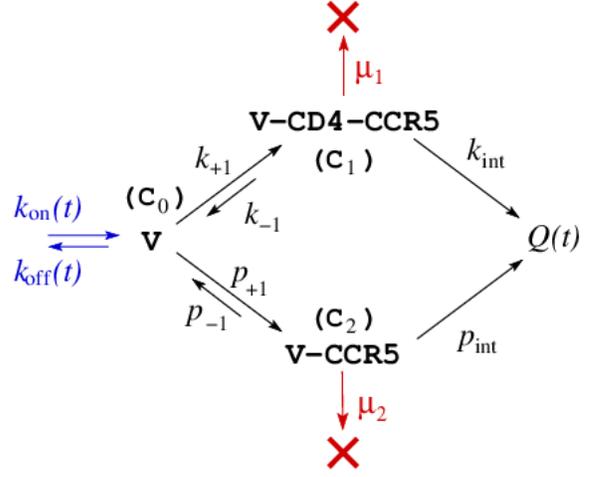}
\caption{Parallel kinetic model of viral entry. In this coarse-grained
  model, V-CD4-CCR5 effectively represents a virus that binds to CD4
  first and is then bound to CCR5. The top pathway effectively
  subsumes the standard sequential entry pathway into a single step.
  Alternatively, we suppose that the virus can interact with CCR5
  directly and also infect the cell with rate $p_{\rm int}$.  This is
  the simplest model that provides a ``CD4-independent'' entry
  pathway.}
\label{PARALLELFIG}
\end{center}
\end{figure}

The quantitative analysis done by Johnston \textit{et
  al.} \cite{Chikere201381, Johnston} showed that HIV-1 infectivity of
some viral strains had remarkably low responsiveness to the induced
surface concentrations of CD4 while still having a relatively
monotonic dependence on CCR5 concentrations, as previously shown in
Fig.~\ref{FIG1}. Furthermore, some strains of simian immunodeficiency
virus (SIV) are known to infect via ``CD4-independent'' pathways,
requiring only CCR5 coreceptor for viral
entry \cite{Edinger1999}. Motivated by these observations, we propose a
``parallel'' pathway model wherein HIV-1 can either enter through the
standard pathway described in the sequential model presented in
the last section, or can either completely bypass CD4 binding. Within
this ``parallel'' model we propose that HIV-1 can interact with CCR5
directly with rate $p_{+1}$ and form the complex V-CCR5 whose
concentration, [V-CCR5], we denote as $c_2(t)$.  As shown in
Fig.~\ref{PARALLELFIG}, this state can then directly enter the cell
through fusion or endocytosis leading to infection with rate $p_{\rm
  int}$ \cite{MELIKIANA}.  For mathematical simplicity, we describe
the standard sequential model discussed in the previous subsection
via a ``lumped'' model where the sequential binding of CD4 and
CCR5 is described by one effective rate.

The corresponding rate equations are:

\begin{eqnarray}
\frac{\dd c_0(t)}{\dd t} &=& k_{\rm{on}}(t) 
-k_{+1}c_0 - p_{+1}c_0 - k_{\rm{off}}c_0  
+ k_{-1}c_1 + p_{-1}c_{2} ,\nonumber \\
\frac{\dd c_1(t)}{\dd t} &=& k_{+1}c_0 - k_{-1}c_1 
- \mu_1 c_1 - k_{\rm int}c_1,\label{PARALLELEQNS} \\
\frac{\dd c_2(t)}{\dd t} &=& p_{+1}c_{0} 
- p_{-1}c_{2}-\mu_{2}c_2 - p_{\rm int}c_2 \nonumber.
\end{eqnarray}
Similar to the sequential model, we expect the binding rates to be
functions of the concentrations of CD4 and CCR5: 
$k_{+1} = k_{+1}^0[\rm{CD4}]^{\beta_1}[\rm{CCR5}]^{\beta_2}$, $p_{+1}
= p_{+1}^0[\rm{CCR5}]^{\gamma_1}$,
where $k_{+1}^0$ and $p_{+1}^0$ are the intrinsic binding rates between
the virus and the respective receptors and $\beta_1$, $\beta_2$, and
$\gamma_{1}$ are effective stoichiometries. The total infectivity is
now given by

\begin{equation}
Q(t) = \int_0^t \left[k_{\rm{int}}c_1(\tau) + p_{\rm{int}}c_2(\tau)\right]\dd\tau.
\label{QPARALLEL}
\end{equation}
To further simplify matters, we set all degradation rates equal so
that $\mu_{1}\approx \mu_{2} = \mu$. Upon solving
Eqs.~\ref{PARALLELEQNS} and \ref{QPARALLEL}, and normalizing by the
reference concentration $Q_{\rm{max}}$ from the infectivity data
associated with $[\mathrm{CD4}]_{\mathrm{max}}$ and
$[\mathrm{CCR5}]_{\mathrm{max}}$, we find the normalized infectivity
to be

\begin{eqnarray}
F_{\infty} & \equiv &\frac{Q_{\infty}}{Q_{\rm{max}}} \nonumber \\
\: & = & {A_{1}Y^{\gamma_{1}y} + A_{2}X^{\beta_{1}x}Y^{\beta_{2}y}\over
  X^{\beta_{1}x}Y^{\beta_{2}y}Y^{\gamma_{1}y}+B_{\rm
    1}Y^{\gamma_{1}y}+B_{\rm 2}X^{\beta_{1}x}Y^{\beta_{2}y}},
    \label{QPARALLELINFTY}
\end{eqnarray}
where 

\begin{align}
A_{1} & ={k_{\mathrm{int}}V_{0} \over Q_{\rm{max}} k_{\mathrm{off}}}{k_{+1}^{0}
\left[\mathrm{CD4}\right]_{\mathrm{min}}^{\beta_1}\left[\mathrm{CCR5}\right]_{\mathrm{min}}^{\beta_2} \over 
\mu+k_{\mathrm{int}}+k_{-1}}, \nonumber \\
A_{2} & ={p_{\mathrm{int}}V_{0} \over Q_{\rm{max}} k_{\mathrm{off}}}{p_{+1}^{0}[\mathrm{CCR5}]_{\mathrm{min}}^{\gamma_{1}}
\over \mu+p_{\mathrm{int}}+p_{-1}},
\label{QPARALLAPAR}
\end{align}
and 

\begin{align}
B_{1} & = {(\mu+k_{\mathrm{int}})k_{+1}^{0}[\mathrm{CD4}]_{\mathrm{min}}^{\beta_1}[\mathrm{CCR5}]_{\mathrm{min}}^{\beta_2} 
\over k_{\mathrm{off}}(\mu+k_{\mathrm{int}}+k_{-1})}, \nonumber \\
B_{2} & = {(\mu+p_{\mathrm{int}})p_{+1}^{0}[\mathrm{CCR5}]_{\mathrm{min}}^{\gamma_{1}} 
\over k_{\mathrm{off}}(\mu+p_{\mathrm{int}}+p_{-1})}.
\label{QPARALLBPAR}
\end{align}

\begin{figure*}[ht]
 \includegraphics[width=6in]{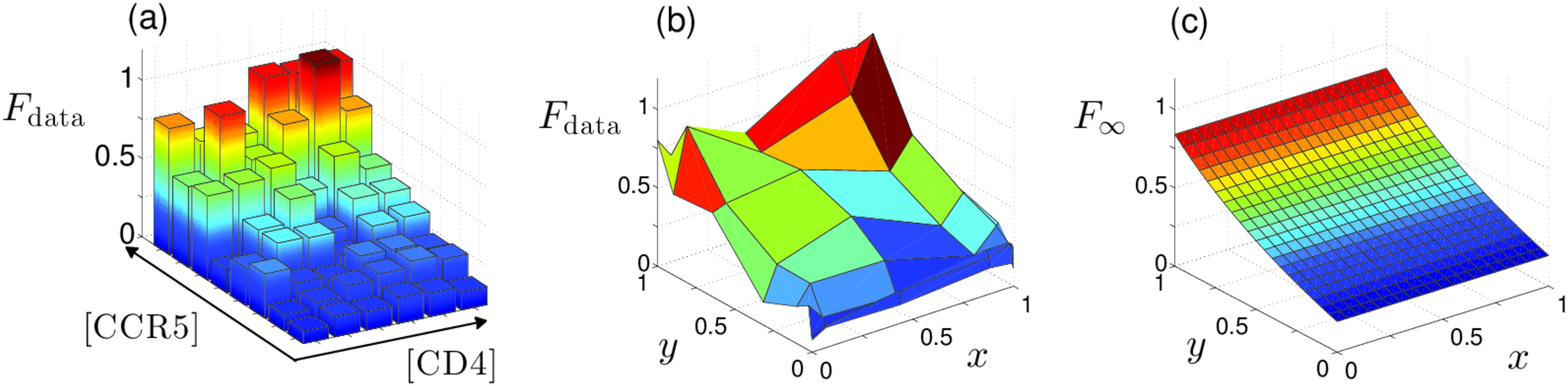}
  \caption{Fitting of the parallel model to the data presented in
    Fig.~\ref{FIG1}, which appears ``CD4-independent.''  (a) Raw
    normalized infectivity data $F_{\mathrm{data}}$ as in
    Fig.~\ref{FIG1}(a).  (b) Scaled normalized data. (c) Best-fit plot
    using maximum likelihood on the parallel model. The maximum
    likelihood parameters are $(\hat{A}_{1}, \hat{A}_{2},
    \hat{B}_{1}, \hat{B}_{2}, \hat{\beta}_{1}, \hat{\beta}_{2},
    \hat{\gamma}_{1}) = (0.23,0.26,1.5,0,0,1.5,2)$ and captures the
    insensitivity to $x$ ([CD4]).}
\label{PARALLELFITS}
\end{figure*}

\noindent
Note that our simplified parallel model has an additional
parameter compared to that of the sequential model. We can now
perform maximum likelihood statistical analysis using our parallel
pathway model. In Fig.~\ref{PARALLELFITS} we show the data of the
NL43(RT) strain from Johnston \textit{et al.} \cite{Johnston}.  The
fitted surface $F_{\infty}(x,y\vert \hat{\boldsymbol\xi})$ shows a
qualitatively good fit to the data. Not surprisingly, $\hat{\beta}_{1}
\approx 0$, indicating the independence of CD4 attachment on [CD4].

In this case, the AIC score yields AIC(parallel) = 3.62 and
AIC(sequential) = -1.3, suggesting that the parallel pathway model
is not statistically warranted even for a highly ``CD4-independent''
strain. Although there are only seven parallel pathway parameters
\textit{versus} six for the sequential model, the parallel
model lumps CD4 and CCR5 binding into a single process which may be
too coarse a description. We test this possibility by exploring this
lumped version of the sequential model.

\begin{figure}[ht]
\begin{center}
\includegraphics[width=3.5in]{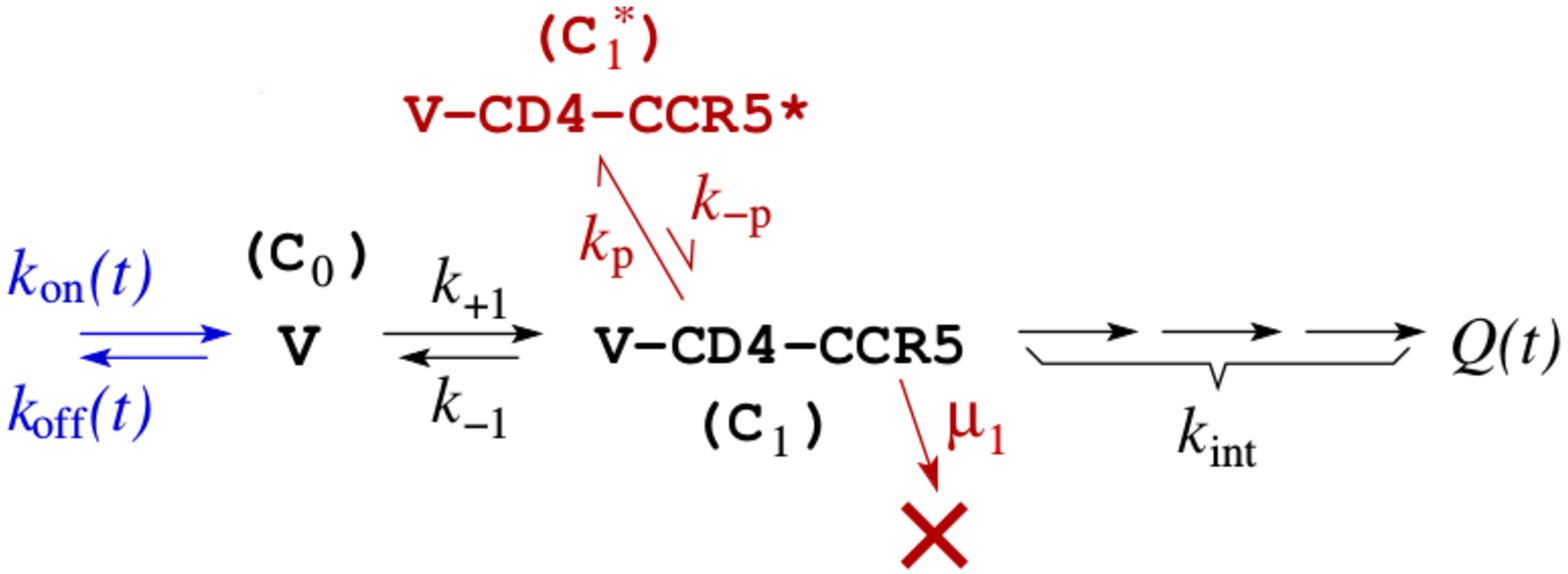}
\caption{Lumped kinetic model of viral entry. The model simplifies
  the sequential model in Fig.~\ref{SERIESFIG} by subsuming the
  intermediate step of HIV-1 binding to CD4 into rate $k_{+1}$ so that
  the virus binds to both CD4 and CCR5 simultaneously.}
\label{LUMPFIG}
\end{center}
\end{figure}

\subsection{Lumped Model}
The sequential model explored above assumes that gp120 binding of
CCR5 is contingent on first binding to the CD4 receptor. Separating
the two viral states V-CD4 and V-CD4-CCR5, which correspond to virus
bound to CD4 and virus bound to both CD4 and CCR5, factors in the
binding and dissociation dynamics between these states into the
expression for infectivity $F_{\infty}$ derived above. If the rate of
transitioning between these two states is sufficiently fast, it is
possible to further simplify the model by eliminating the intermediate
state V-CD4 by assuming that CD4 and CCR5 binding occur
simultaneously, as shown in Fig.~\ref{LUMPFIG}. Upon simplifying the
model in this manner, we reduce the parameter space for which we
perform statistical inference. In order to explore whether this
simplification leads to a better model fit and estimation of physical
rate parameters, we start with the rate equations:
\begin{eqnarray}
\frac{\dd c_0(t)}{\dd t} &=& -k_{+1}c_0 - k_{\rm{off}}c_0 + k_{-1}c_1,
\nonumber \\ 
\frac{\dd c_1(t)}{\dd t} &=& k_{+1}c_0 - k_{-1}c_1 - \mu_1c_1 -
k_{\rm{p}}c_1 - k_{\rm{int}}c_1 + k_{-\rm{p}}c_{1}^*, \nonumber \\
\frac{\dd c_{1}^{*}(t)}{\dd t} &=& k_{\rm{p}} c_1 - k_{-\rm{p}}c_{1}^{*} -
\mu_{\rm p} c_{1}^{*}.
\label{LUMPEQN}
\end{eqnarray}

Within the the lumped model, we expect the rate of simultaneous
CD4 and CCR5 binding to be a function of both the concentrations of
CD4 and CCR5: $k_{+1} =
k_{+1}^0[\rm{CD4}]^{\beta_1}[\rm{CCR5}]^{\beta_2}$, where $\beta_1$
and $\beta_2$ are the appropriate stoichiometry parameters, similar to
those defined in the sequential model. Here, the raw infectivity
is $Q(t) =k_{\rm{int}} \int_0^t c_1(\tau) \dd\tau$ while the
normalized rescaled infectivity takes the form
\begin{equation}
F_{\infty} \equiv \frac{Q_{\infty}}{Q_{\rm{max}}} = \frac{C_2}{1 + C_1X^{\beta_1x}Y^{\beta_2y}},
    \label{QLUMPINFTY}
\end{equation}
where
\begin{eqnarray}
C_1 &=& \left(\frac{1}{[\rm{CD4}]_{\rm{min}}}\right)^{\beta_{1}}
\left(\frac{1}{[\rm{CCR5}]_{\rm{min}}}\right)^{\beta_{2}}
\left(\frac{k_{\rm{off}}\omega}{k_{+1}^0\left(\omega-k_{-1}\right)}\right), \nonumber \\
C_2 &=& \frac{k_{\rm{int}}V_0}{Q_{\rm{max}} \left(\omega - k_{-1}\right)}.
\label{LUMPAB}
\end{eqnarray}
As in the sequential model, the quantity
\begin{equation}
\omega = k_{-1} + k_{\rm{int}} + \left(\frac{\mu_{\rm{p}}}{k_{\rm{-p}}+\mu_{\rm{p}}}\right)k_{\rm{p}} 
+ \mu_1
\end{equation}
can be considered the bulk flow of virus out of the V-CD4-CCR5 viral state. 

\begin{figure*}[ht]
\begin{center}
\includegraphics[width=6in]{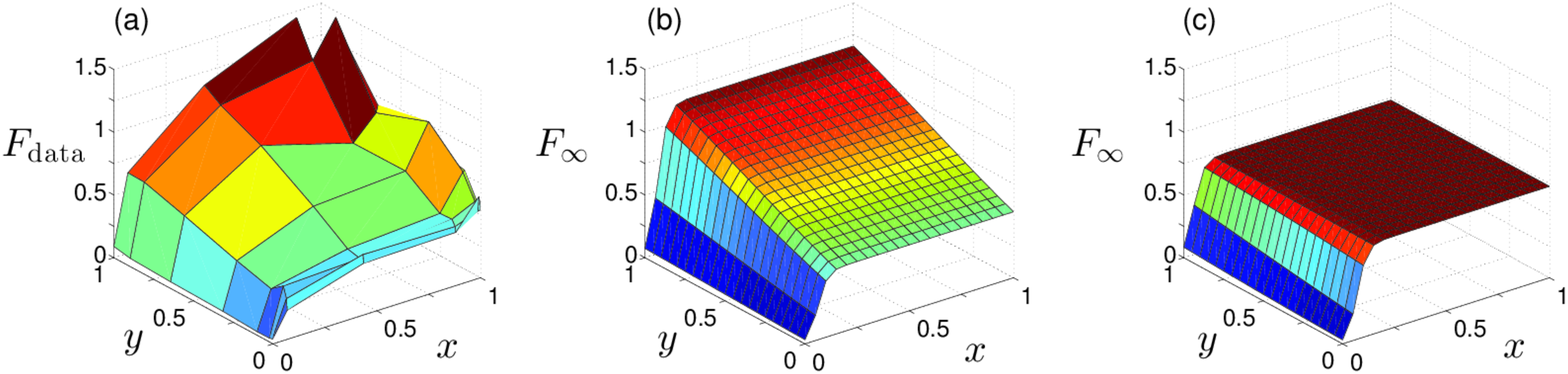}
\caption{A comparison between the lumped model and the
  sequential model. (a) Scaled normalized data
  $F_{\mathrm{data}}$. (b) Best fitted plot using maximum
  likelihood of the sequential model from
  Fig.~\ref{SERIESFIG}. (c) Best fitted plot using maximum
  likelihood of the lumped model with estimated parameters
  $(\hat{C}_{1}, \hat{C}_{2}, \hat{\beta}_{1}, \hat{\beta}_{2}) =
  (22.0,0.73,10.6,0.37)$. The lumped model's functional form, compared
  to that of the sequential model, prevents a qualitatively accurate
  representation of the data, especially in the $y$-dependence of
  $F_{\rm data}$.}
\label{LUMPFITS}
\end{center}
\end{figure*}

Upon comparing AIC scores of AIC(sequential) = 15.6 and AIC(lumped) =
17.6 from the viral strain B5 (YA) data presented in Johnston
\textit{et al.} \cite{Johnston}, we find the sequential model
yields a better fit to the data, despite the reduction of the number
of parameters in the lumped model.  Fits are shown in
Fig.~\ref{LUMPFITS}.  As in the sequential model, $\beta_1$
dictates the sharpness of the descent of $F_{\infty}$ for very low
values of [CD4], but the terms that define the tilt of the broader
region of the function are lost in the lumped model, preventing an
adequate fit of the average slope of the data. This signifies that the
intermediate process of CD4 binding prior to CCR5 binding is a
necessary inclusion into the model.


\section{Conclusions}

In order to distinguish different strains of HIV-1 via their entry
kinetics, heuristic metrics have been derived to classify different
sensitivities of infection to CD4 and CCR5 expression in the host cell
\cite{Johnston}.  The metrics $M$, $\Delta$, and $\theta$ in
Eqs.~\ref{OLDMETRICS} are good discriminators and cluster experimental
replicates of identical strains of HIV-1 sufficiently due to the fact
that they are purely based on the shape of the data, but not on any
mechanistic processes involved in viral entry. Here, we analyzed a
sequential kinetic binding model of HIV-1 viral entry that yields
a functional relationship between the infectivity of a strain of virus
and the levels of CD4 and CCR5 based on known chemical processes.  Our
model provides a framework in which to analyze different strains of
HIV-1 based on combinations of parameters in the kinetic model and
physical insight into how these parameters facilitate or inhibit HIV-1
viral entry. One can now distinguish different strains of HIV-1
according to the inferred values of kinetic parameters and display the
infectivity of each strain as points in parameter space with physical
meaning.

In addition to kinetic rates, stoichiometries of CD4 and CCR5 are
incorporated in our model. In fact, the dependencies of the
infectivity to CD4 and CCR5 levels are most sensitive to their
respective stoichiometries at the expense of the sensitivity of the
other estimated parameters. We show that overall infectivity data is
sufficient to provide some confidence in assigning nonunit
stoichiometries, suggesting that on average either multiple CD4s bind
to a gp spike, or multiple gp spikes are engaged in a typical entry
event. In fact, due to the high sensitivity of the characteristic
shape of the infectivity function to stoichiometry, we suggest that
experiments be designed to increase the number of data points in
regions of high gradients of $F_{\rm data}(x,y)$, where the
stoichiometric parameters $\beta_{1,2}$ exhibit the most influence. In
this same regard, choosing the minimum and maximum values for the CD4
and CCR5 experimental inputs dictates the relative sizes of the
constants $X$ and $Y$, thus altering the relative sensitivity between
the two stoichiometric parameters. These properties of
$F_{\infty}(x,y)$ provide guidance to the experimentalist in designing
the most informative measurements.

%
%


Finally, in order to address the existence of strains that are highly
insensitive to CD4 expression, and that infect cells with extremely
low levels of CD4, we proposed a parallel pathway model that
allows slow entry, even in the absence of CD4, through, perhaps, an
endocytotic mechanism \cite{MELIKIANA}. Here, binding to CCR5 is
sufficient to allow for viral entry through an alternate pathway. We
performed parameter inference on this parallel model and compared
our results with those from the original sequential model. We also
explored a simplification of the sequential model by subsuming the
intermediate CD4 binding process into one combined process of
simultaneous CD4 and CCR5 binding. We compared the performance of this
last lumped model with the sequential model to determine what
effects such simplifying assumptions might have on the inference
capabilities of our models.

The modeling and data analysis framework we developed in this work may
also be used to quantify the effectiveness of fusion inhibitors. For
example, though fusion inhibitors can arrest the fusion process at an
intermediate step \cite{MELIKIANB}, the bond between gp120 and gp41 is
non-covalent and weak enough so there is a high probability of this
bond breaking, resulting in the virus dissociating from the
cell \cite{Manca01091993}. Though the Env spike used in that failed
infection attempt is now non-functional, the virus can theoretically
return to the cell and make another attempt with a different Env on
its membrane. But unlike simian immunodeficiency virus (SIV) that is
covered in large amounts of spikes \cite{Compans}, HIV-1 has relatively
small numbers of Env on its surface; often on the order of five spikes
per virus \cite{Brown04031997}. Thus fusion inhibition becomes a
process of repeated failed attempts at infection of a virus until the
Env spikes are depleted. In this context, two time scales would be
established between the rate of fusion without inhibitor and the rate
of depletion of glycoprotein spikes. Modeling this aspect of the
infectivity process in the presence of fusion inhibitors can give
better insight into the effectiveness of inhibitor treatment and
recommendations on the duration of possible treatment protocols. As
shown in Fig.~\ref{QT}(b), a model of infectivity can be used to
predict a reduction in viral infection as a function of fusion
inhibitor dosage.  Instead of performing infectivity measurements for
varying [CD4] and [CCR5], we can also change fusion inhibitor levels
and study the corresponding infectivity patterns using the same models
presented in this work.  Similar analyses can also be performed to
study the efficacy of broadly neutralizing antibodies in suppressing
viral entry.  These and other physical chemical considerations will be
pursued in future work involving model analysis and inference from
forthcoming data.

\section{acknowledgement}
The authors thank Paul Gorry for sharing some of infectivity data sets
presented in this work.  BM was supported by the NIH through a T32
training grant GM008185. TC acknowledges support from the National
Science Foundation through grant DMS-1516675. MD and TC were also
supported by Army Research Office through grant W911NF-14-1-0472.

%
%


\bibliography{REFS}

\end{document}